\documentclass[conference]{IEEEtran}
\IEEEoverridecommandlockouts
\usepackage{cite}
\usepackage{hyperref} 
\usepackage{amsmath,amssymb,amsfonts}
\usepackage{algorithmic}
\usepackage{graphicx}
\usepackage{textcomp}
\usepackage{xcolor}
\usepackage{graphicx}
\usepackage{array}
\usepackage{booktabs}
\usepackage{enumitem}
\usepackage{colortbl}
\usepackage{xcolor}
\usepackage{siunitx}
\usepackage[linesnumbered,ruled,vlined]{algorithm2e}
\usepackage{graphicx} 
\usepackage{multirow}
\usepackage{subcaption}

\usepackage{newtxtext,newtxmath}
\def\BibTeX{{\rm B\kern-.05em{\sc i\kern-.025em b}\kern-.08em
    T\kern-.1667em\lower.7ex\hbox{E}\kern-.125emX}}

\begin{document}

\title{Context-Aware Information Transfer via Digital Semantic Communication in UAV-Based Networks}

\author{\IEEEauthorblockN{$^1$ Poorvi Joshi, $^2$Mohan Gurusamy}
\IEEEauthorblockA{ \textit{Department of Electrical and Computer Engineering}\\ \textit{National University of Singapore, Singapore}\\
 Email: $^1$e1144005@u.nus.edu, $^2$gmohan@nus.edu.sg}}



\maketitle

\begin{abstract}
In smart cities, bandwidth-constrained Unmanned Aerial Vehicles (UAVs) often fail to relay mission-critical data in time, compromising real-time decision-making.  This highlights the need for faster and more efficient transmission of only the most relevant information. To address this, we propose DSC-UAV model, leveraging a context-adaptive Digital Semantic Communication (DSC) framework. This model redefines aerial data transmission through three core components: prompt-aware encoding, dynamic UAV-enabled relaying, and user mobility-optimized reinforcement learning. Ground users transmit context-driven visual content. Images are encoded via Vision Transformer combined with a prompt-text encoder to generate semantic features based on the desired context (generic or object-specific). These features are then quantized and transmitted over a UAV network that dynamically relays the data. Joint trajectory and resource allocation are optimized using Truncated Quantile Critic (TQC)-aided reinforcement learning technique, which offers greater stability and precision over standard SAC and TD3 due to its resistance to overestimation bias. Simulations demonstrate significant performance improvement, up to 22\% gain in semantic-structural similarity and 14\% reduction in Age of Information (AoI) compared to digital and prior UAV-semantic communication baselines. By integrating mobility control with context-driven visual abstraction, DSC-UAV advances resilient, information-centric surveillance for next-generation UAV networks in bandwidth-constrained environments.

\end{abstract}


\section{Introduction}

In smart city surveillance systems, the transmission of high-resolution images and video frames plays a pivotal role in real-time monitoring and threat detection \cite{10530547}. To enhance connectivity with centralized servers and ensure low-latency delivery, Unmanned Aerial Vehicles (UAVs) have emerged as a promising solution. Due to their mobility, flexibility, and ability to function as mobile edge computing (MEC) servers, UAVs enable faster and more efficient data collection and transmission \cite{10530547,10922390}. However, the exponential increase in surveillance data, combined with the growing density of interconnected smart devices and vehicular networks, has imposed severe demands on wireless infrastructure \cite{10922390}. In this context, traditional UAV-based communication systems face two critical challenges: limited bandwidth availability and constrained onboard energy resources \cite{10044272, 10930657}. These limitations not only hinder real-time data transmission but also restrict the operational lifespan and coverage capabilities of UAVs in large-scale surveillance deployments \cite{10930657}. 

To address these limitations, semantic communication has recently gained significant attention. Unlike conventional methods that transmit raw bit-level data, semantic communication focuses on conveying task-relevant or intent-level information \cite{9955312}. In the context of UAV-assisted networks, recent approaches have demonstrated substantial bandwidth savings by transmitting only high-level semantic features rather than full-resolution sensor data. For example, the PE-MMSC framework \cite{guo2025perception} fuses hyperspectral and LiDAR semantics onboard UAVs to reduce transmission volume while maintaining classification accuracy under low-SNR conditions. Similarly, VAE-based encoders have been employed to extract latent semantic representations from UAV imagery, enabling efficient transmission of compressed features that preserve semantic similarity \cite{10437643}. These methods significantly reduce bandwidth consumption while ensuring high reconstruction fidelity and robust task performance for downstream applications such as object detection and scene understanding. While these methods reduce bandwidth consumption and maintain high reconstruction fidelity, they face two key limitations. First, they transmit analog semantic features directly, which makes them highly susceptible to channel noise and difficult to integrate with digital hardware, thereby necessitating digital encoding. Second, they lack context-awareness, which is critical for intelligent surveillance tasks. \textit{For instance, if the task is to focus on red cars at high resolution in a congested traffic scenario, context-unaware models may indiscriminately extract all scene elements—such as traffic lights, bicycles, and unrelated vehicles—at lower resolutions, thereby compromising task relevance and overall system efficiency.}


To address the limitations of prior UAV-assisted semantic communication systems—includes, the lack of context-awareness, digital quantization, and joint optimization—we propose a Context-Aware Digital Semantic Communication for UAV network (DSC-UAV) framework for mobile surveillance networks. Each Ground User (GU) (e.g., dashcam-equipped vehicles or fixed CCTVs) is served by multiple UAVs acting as parallel relays, enabling faster and more reliable data transfer. Our main contributions are: 

\begin{itemize}
    \item We propose a prompt-aware semantic encoder-decoder that fuses visual features from a Vision Transformer (ViT) with task-specific text prompts. The joint representation is processed through a sparse neural network to extract compressed semantic features. At the receiver, the same prompt guides a CNN-based decoder for accurate and context-driven image reconstruction, enabling intent-aware and efficient communication.
    \item We develop a Truncated Quantile Critic (TQC)-based reinforcement learning (RL)method to jointly optimize UAV trajectories, compression ratios, and relay-task allocation for parallel UAV relaying. The approach targets minimizing Age of Information (AoI) and maximizing min Semantic Structural Similarity (SSS), achieving stable learning and enhanced performance in dynamic network conditions.
    \item We evaluate the framework on two surveillance scenarios: generic scene understanding and object-specific intent. Our experiments show up to 14\% reduction in AoI and 22\% improvement in SSS compared to baselines. We also analyze the effect of semantic codeword length, modulation schemes, and update intervals on system performance.
\end{itemize}

\section{Related Works}

\subsection{Digital Semantic Communication}
Quantization plays a pivotal role in digital semantic communication (DSC) by bridging continuous semantic representations and discrete digital signals. Various approaches such as scalar quantization, vector quantization (VQ), and non-linear adaptive schemes have been explored to achieve compact and compatible encodings for digital transmission \cite{zhang2024towards}. Among these, VQ is widely adopted due to its effectiveness in converting high-dimensional semantic embeddings into discrete codewords. This reduces transmission overhead and  facilitates integration with digital modulation schemes \cite{zhang2024towards}. In \cite{10065571}, VQ-DeepSC incorporates multi-scale embedding with hard quantization via nearest-neighbor search, achieving robust performance under noisy conditions.

However, hard quantization introduces non-differentiability, which affects end-to-end learning, a critical requirement in our system where partial joint optimization of the semantic encoder, UAV network parameters, and decoder is done. To address this, we employ a soft-to-hard quantization strategy, which begins with soft assignments and progressively anneals them into discrete representations. This enables differentiable training while preserving the final discretization necessary for digital transmission. A representative work in this direction is \cite{agustsson2017soft}, which proposed soft-to-hard vector quantization for compressible deep representations, demonstrating stable training dynamics and strong compression performance. These quantization approaches, however, have not been fully adapted to UAV communication scenarios where bandwidth and energy limitations amplify the need for quantization-aware design.

\subsection{Semantic Communication in UAV Network}

Semantic communication (SemCom) has gained importance for improving UAV-assisted network efficiency under energy and bandwidth constraints. In \cite{10437643}, a hybrid-action deep RL framework is proposed that jointly optimizes UAV trajectory, transmit power, and semantic model scaling, balancing reconstruction quality with computational energy cost. Zhao et al.\cite{10255282} developed a scene graph-based semantic encoder combined with a combinatorial auction-based relay selection mechanism for metaverse data delivery, enhancing semantic richness and content freshness. In \cite{10527365}, a semantic entropy-guided relay selection strategy is introduced, integrating an energy-aware incentive mechanism to balance semantic entropy gain against UAV energy efficiency. Song et al. \cite{10872947} presented a multi-scale semantic encoder using knowledge graph-based feature extraction to improve UAV-assisted object detection by reducing semantic distortion and transmission overhead.

Despite these advances, prior works overlook digital quantization of semantic features and rely on energy-intensive onboard UAV decoding, limiting practical deployment in resource-constrained networks. Our DSC-UAV framework addresses these gaps by modeling UAVs as relay nodes for parallel transmission, offloading decoding to a central server to reduce computational overhead. We employ quantization to enhance robustness towards channel noise and bandwidth efficiency. Our approach optimizes data freshness (AoI) and Semantic-Structural Similarity (SSS), accounting for both semantic similarity and visual fidelity.

\subsection{Task Oriented Semantic Communication}

Task-oriented semantic communication (TSC) addresses the limitations of conventional semantic systems by extracting and transmitting only task-relevant features, improving bandwidth efficiency and task performance. In \cite{9653664}, Transformer-based task-specific encoding is demonstrated, which significantly improves downstream accuracy in image retrieval, machine translation, and visual question answering. Fu et al. \cite{10226153} developed an attention-driven architecture supporting image reconstruction and object detection, selectively emphasizing task-critical spatial features for higher fidelity. Ma et al. \cite{10110357} proposed a $\beta$-VAE-based framework for interpretable semantic feature selection, achieving robust task performance under semantic noise.
Building on these advances, our design uses a ViT encoder guided by CLIP-based textual prompts to extract context-aware features, compressed via Sparse NN for bandwidth efficiency, and decoded using a CNN-based decoder to reconstruct spatially consistent outputs.

\section{System Model}

\begin{figure}[t]
    \centering
    \includegraphics[width=1.0\linewidth]{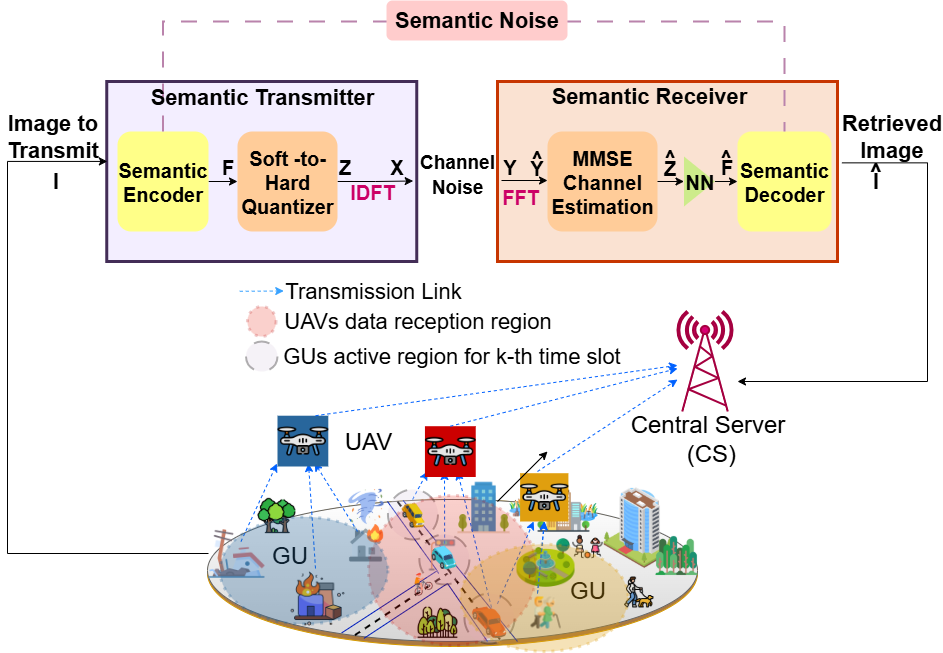} 
    \caption{System Model}
    \label{fig:SM}
\end{figure}

In this study, we define a UAV-aided mobile network that builds on context-aware data transmission under limited bandwidth but with higher transmission efficiency, as shown in Fig.~\ref{fig:SM}. In this system, we consider $M$ GUs denoted by $\{1,2,\dots,m,\dots,M\}$ and $N$ UAVs denoted by $\{1,2,\dots,n,\dots,N\}$, along with a central server located at the origin. Each GU periodically transmits real-world images to the central server with an image arrival rate of $\lambda_m$. The GUsare equipped with a semantic transmitter, which encodes the images before transmission, while the central server employs a semantic receiver to decode them. The data transmission is carried over an orthogonal frequency division multiplexing (OFDM) system to enable efficient parallel communication. In this network, the UAVs act as relays, providing parallel transmission paths to enable faster data updates. A single GU can be served by multiple UAVs simultaneously, thereby increasing transmission diversity. Importantly, data transmission from a GU begins only after all its assigned UAVs have reached their optimal locations for reliable and efficient communication. The wireless channel in the system follow the Nakagami-$m$ distribution \cite{1412063}.

\subsection{UAV State and Energy Consumption}

\begin{figure}[ht]
    \centering
    \includegraphics[width=1\linewidth]{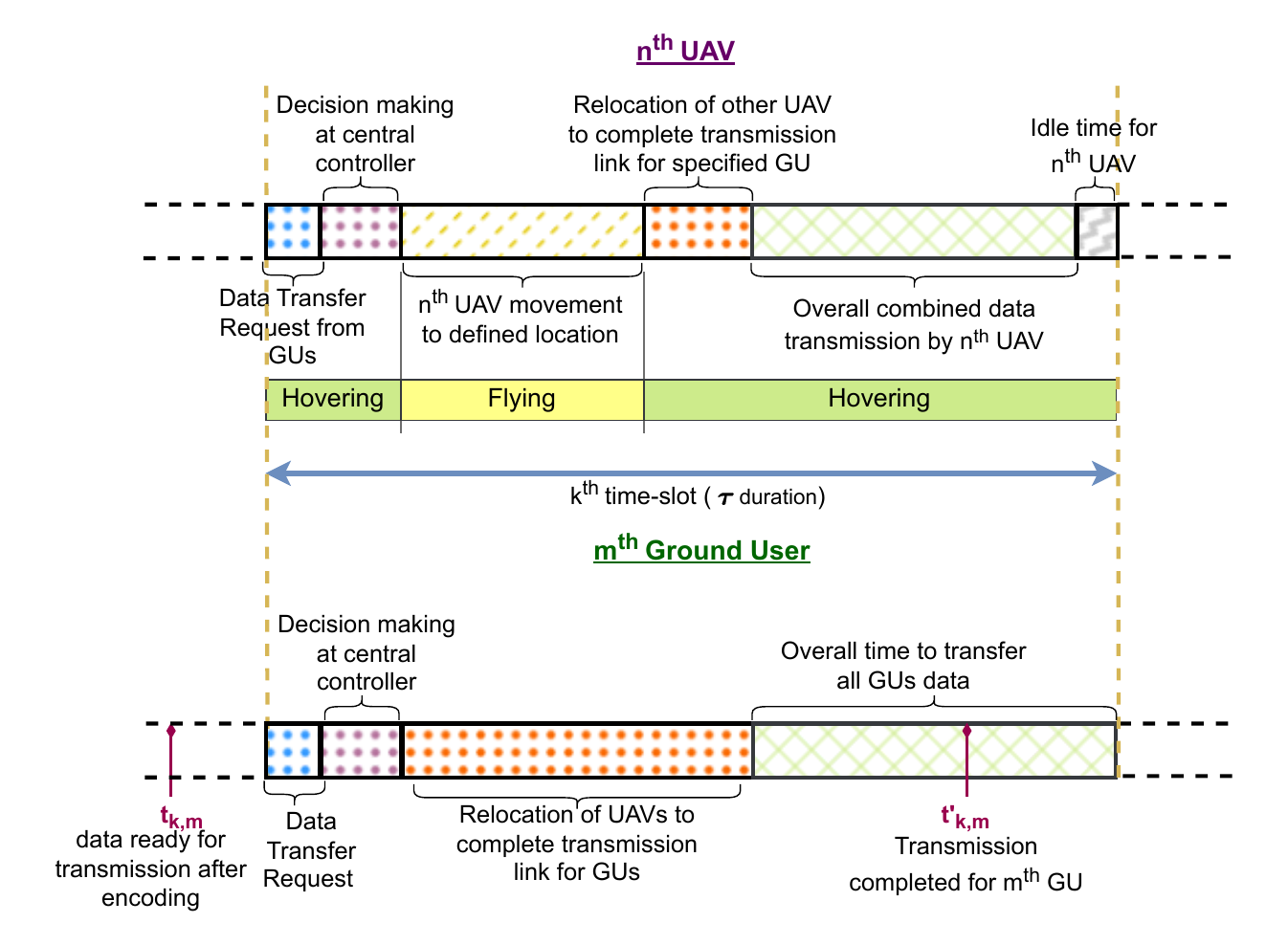} 
    \caption{$k^{th}$ Time Slot for $n^{th}$ UAV and $m^{th}$ GU}
    \label{fig:TS}
\end{figure}

The mission time $T$ with timestamp $t$ is divided into $K$ time slots, each of duration $\tau \leq \min \{\lambda_m\}$, ensuring that in each time slot at most one image data from each GU is addressed by the UAV network. Fig. ~\ref{fig:TS} illustrates the $k^{th}$ time slot from both the UAV and GU perspectives. At the beginning of each time slot, a data request transfer occurs. Each GU $m$ transmits its current position $ \mathbf{s}^{\mathrm{GU}}_m(t) = [x^{\mathrm{GU}}_m(t), y^{\mathrm{GU}}_m(t), 0]$ at $t = k\tau$, along with its current speed $v^{\mathrm{GU}}_m(t)$. We assume that within the given time slot $t \in [(k-1)\tau, k\tau]$, the speed remains constant and is denoted by $v^{\mathrm{GU}}_{m,k}$. In addition, the GUs share the time at which the encoded data will be ready for transmission $t_{k,m}$ and the data size to be transmitted $D_m(k)$. After this, the central server performs decision-making to determine the UAVs' desired locations. The UAVs then start relocating to their target positions. Once all UAVs have reached their optimal locations, data transmission begins. When the data transmission for the $m^{th}$ GU is completed, the completion timestamp $t'_{k,m}$ is recorded. The task for the $k^{th}$ time slot ends when the transmission of all GUs' data is complete. We assume that the data transfer request time and the decision-making time are comparatively smaller than the UAV flying time and the communication time. Therefore, these two delays are neglected in our system analysis. The $n^{th}$ UAV will operate in flying mode during its relocation, and the rest will be in hover mode.


The position of the $n^{th}$ UAV at timestamp $t$ is expressed as
$
\mathbf{s}^{\mathrm{UAV}}_n(t) = [x^{\mathrm{UAV}}_n(t), y^{\mathrm{UAV}}_n(t), z^{\mathrm{UAV}}_n(t)].
$
Each UAV moves with speed $v^{\mathrm{UAV}}_n(t)$, covering a distance 
$
l^{\mathrm{UAV}}_n(t) = v^{\mathrm{UAV}}_n(t) \cdot 1
$
in unit time, in the direction of the angular vector 
$
\hat{\omega}_n(t) = \{\omega_{n}^{\mathrm{el}}(t), \omega_{n}^{\mathrm{az}}(t)\}.
$
Here, $\omega_{n}^{\mathrm{el}}(t) \in [0, \pi]$ is the elevation angle, and $\omega_{n}^{\mathrm{az}}(t) \in [0,2\pi)$ is the azimuthal angle in the horizontal plane. After relocation, the position of the $n^{th}$ UAV for the $k^{th}$ time slot is denoted by $\mathbf{s}^{\mathrm{UAV}}_{n,k}$. Assuming that all UAVs have the same coverage angle $\alpha_r$, the horizontal radius of the data-receiving region for the $n^{th}$ UAV during time slot $k$ is given by
$
C^{\mathrm{max}}_{n,k} = z^{\mathrm{UAV}}_{n,k} \tan(\alpha_r).
$
Therefore, the data-receiving region of the $n^{th}$ UAV in time slot $k$ is defined as,
\begin{equation}\label{DRR}
R^{n}_{\mathrm{dr},k} = \{(x,y) : (x - x^{\mathrm{UAV}}_{n,k})^2 + (y - y^{\mathrm{UAV}}_{n,k})^2 \leq (C^{\mathrm{max}}_{n,k})^2\}.
\end{equation}


Finally, each UAV has a finite energy budget $E_{\max}$ that must be considered in its operations. Within slot $k$, UAV $n$ spends $\tau^{n}_{k,\mathrm{move}}$ seconds in relocation and the remaining time hovering. The propulsion energy consumed due to flight state is modeled as
\begin{equation}
E^{\mathrm{St}}_{n}(k) = \alpha_{\mathrm{move}} \, \tau^{n}_{k,\mathrm{move}} + \alpha_{\mathrm{hover}} \, \big(\tau - \tau^{n}_{k,\mathrm{move}}\big),
\end{equation}
where $\alpha_{\mathrm{move}}$ and $\alpha_{\mathrm{hover}}$ are the average propulsion power coefficients (watts) for movement and hovering. These coefficients are evaluated using the rotary-wing UAV power model \cite{mozaffari2019tutorial}:
\begin{equation}
\alpha(v) = c_1 \left( 1 + \frac{3v^2}{v_{\mathrm{tip}}^2} \right)
+ c_2 \left( \sqrt{1 + \frac{v^4}{4v_0^4} - \frac{v^2}{2v_0^2}} \right)
+ \frac{1}{2} c_3 v^3,
\end{equation}
with $\alpha_{\mathrm{move}} = \alpha(v^{\mathrm{UAV}}_{n,k})$ and $\alpha_{\mathrm{hover}} = \alpha(0)$. Here, $v^{\mathrm{UAV}}_{n,k}$ is the average flight speed during relocation, $v_{\mathrm{tip}}$ is the rotor blade tip speed, $v_0$ is the induced velocity in hover, and $c_1,c_2,c_3$ are constants related to UAV power, rotor geometry, and air density. The relocation time will be given as,
\begin{equation}
\tau^{n}_{k,\mathrm{move}} = \min\left\{ \frac{ \displaystyle \int_{(k-1)\tau}^{k\tau} l^{\mathrm{UAV}}_n(t) \, dt }{ v^{\mathrm{UAV}}_{n,k} }, \quad \tau \right\}.
\end{equation}

\subsection{Data Transmission and Reception}

Each GU $m$ in time slot $k$ is equipped with a semantic transmitter that processes the transmitting image
$
I_{m}(k) \in \{0,1,2, \ldots, 255\}^{C \times H \times W},
$
where $C$, $H$, and $W$ denote the number of channels, height, and width of the image, respectively. The image is processed by a semantic encoder $V_{\theta}(\cdot)$ to produce the semantic feature representation
$$
F_{m}(k) = V_{\theta}(I_{m}(k)),
\qquad
F_{m}(k) \in \mathbb{C}^{C \times \frac{HW}{2^{2d}}}
$$
where $d$ is the compression factor.  
Next, the semantic feature is passed through a soft-to-hard quantizer $Q(\cdot)$, which is implemented using a self-attention module inspired by~\cite{10851324}, to generate
$
Z_{m}(k) = Q(F_{m}(k)).
$
We then apply the inverse discrete Fourier transform (IDFT) to obtain the time-domain transmitting symbols:
$$
x_{m}(k) = \mathrm{IDFT}(Z_{m}(k)),
\qquad
x_{m}(k) \in \mathbb{C}^{C \times \frac{H W}{2^{2d}}}
$$

The size of the data to be transmitted for GU $m$ in time slot $k$ can be expressed as
\begin{equation}
D_{m}(k) = D_{\mathrm{orig}} \times \frac{1}{2^{2d}},
\end{equation}
where the original image size in bits is 
$
D_{\mathrm{orig}} = 8 C H W.
$  

We employ OFDM for wireless transmission, where each GU--UAV link $(m,n)$ in slot $k$ is assigned a dedicated set of subcarriers. At the UAV receiver $n$, after cyclic-prefix removal and FFT, minimum mean square error (MMSE) channel estimation is performed. The equalized received frequency-domain symbols are denoted as
$
\widehat{Z}_{m,n}(k).
$
Using the real and imaginary components of $\widehat{Z}_{m,n}(k)$, the estimated semantic feature $\widehat{F}_{m}(k)$ is reconstructed using the inverse mapping of the quantizer. Finally, the semantic decoder $C_{\phi}(\cdot)$ at the central server reconstructs the image:
$$
\widehat{I}_{m}(k) = C_{\phi}\big(\widehat{F}_{m}(k)\big).
$$

The semantic encoder $V_{\theta}(\cdot)$ is based on a ViT, while the semantic decoder $C_{\phi}(\cdot)$ is a CNN-based model that incorporates a text-prompt encoder for improved context awareness. In subsequent sections, we focus on the encoder and decoder architectures.

\subsection{Transmission Protocol}

The transmission of data from the $m^{th}$ GU via the $n^{th}$ UAV in the $k^{th}$ time slot will only occur if the region of the $m^{th}$ GU, $R^{m}_{\mathrm{GU},k}$, lies within the data-receiving region of the $n^{th}$ UAV, $R^{n}_{\mathrm{dr},k}$, as defined in Eq.~\ref{DRR}. The region of the $m^{th}$ GU during the $k^{th}$ time slot is expressed as
\begin{equation}
R^{m}_{\mathrm{GU},k} = \{(x,y) : (x - x^{\mathrm{GU}}_{m,k})^2 + (y - y^{\mathrm{GU}}_{m,k})^2 \leq (v^{\mathrm{GU}}_{m,k} \cdot \tau)^2\},
\end{equation}
where $\mathbf{s}^{\mathrm{GU}}_{m,k}$ denotes the position of the $m^{th}$ GU at the beginning of the time slot $t = (k-1)\tau$. A single GU can be served by multiple UAVs simultaneously. We define the binary variable $\delta^{n}_{m}(k) \in \{0,1\}$ to indicate whether the $m^{th}$ GU is served by the $n^{th}$ UAV in the $k^{th}$ time slot. If $\delta^{n}_{m}(k) = 1$, it implies that the $m^{th}$ GU is actively being served by the $n^{th}$ UAV. The total number of UAVs serving each GU satisfies
$$0 < \sum_{n=1}^{N} \delta^{n}_{m}(k) \leq N, \quad \forall m.$$ 
We further define $\rho^{n}_{m}(k) \in [0,1]$ as the fraction of the $m^{th}$ GU's data transmitted via the $n^{th}$ UAV during the $k^{th}$ time slot. The data allocation across UAVs satisfies the condition
$$\sum_{n=1}^{N} \rho^{n}_{m}(k) \leq 1, \quad \forall m.
$$ 

\subsubsection{G2A Transmission}

In the G2A (Ground-to-Air) transmission phase, the $m^{th}$ GU transmits $D_m(k)$ data bits in the $k^{th}$ time slot to the $n^{th}$ UAV. The received signal at UAV $n$ can be expressed as  
\begin{equation}
    y^{\mathrm{UAV}}_n(k) = 
    h_{m,n}^{GU} 
    \big( d^{\mathrm{GU}}_{m,n}(k) \big)^{-p}
    e^{j 2\pi f_D(k).\tau} 
    x_{m,n}(k) 
    + n^{\mathrm{UAV}}_n,
    \label{eq:g2a}
\end{equation}
where $h_{m,n}^{GU}$ denotes the channel coefficient between GU $m$ and UAV $n$ that follows a Nakagami-$m$ fading distribution, and $d^{\mathrm{GU}}_{m,n}(k)$ is the distance between GU $m$ and UAV $n$ during the $k^{th}$ time slot. The term $p$ represents the path loss exponent, while $f_D(k)$ accounts for the Doppler frequency shift due to GU mobility and is given by  
\begin{equation}
    f_D(k) = \frac{v^{\mathrm{GU}}_m(k) f_c \cos(\theta_{m,n}(k))}{c},
    \label{eq:doppler}
\end{equation}
where $v^{\mathrm{GU}}_m(k)$ is the speed of GU $m$, $f_c$ is the carrier frequency, $\theta_{m,n}(k)$ is the angle between the GU's velocity vector and the line connecting GU $m$ and UAV $n$, and $c$ is the speed of light. The signal $x_{m,n}(k)$ represents the portion of the encoded data symbols of GU $m$ transmitted with unit power to UAV $n$ in the $k^{th}$ time slot based on the allocation ratio $\rho^n_m(k)$, and $n^{\mathrm{UAV}}_n(k)$ denotes the complex additive white Gaussian noise at the UAV receiver, modeled as $ n^{\mathrm{UAV}}_n(k) \sim \mathcal{CN}(0, {\sigma^{UAV}_n}^2).$

\subsubsection{A2G Transmission}

In the A2G (Air-to-Ground) phase, UAV $n$ forwards—via amplify-and-forward (AF)—the signal it received from GU $m$ in slot $k$. A per-stream power $\frac{P^{\mathrm{UAV}}}{M_n(k)}$ is allocated, where $M_n(k)=\sum_{m=1}^{M}\delta_m^n(k)$ is the number of GUs served by UAV $n$ in slot $k$. The received signal at the central server is
\begin{equation}
    y_{m,n}^{\mathrm{CS}}(k) = h_n^{\mathrm{CS}}\, G_{m,n}(k)\, y_n^{\mathrm{UAV}}(k) + n^{\mathrm{CS}},
    \label{eq:a2g}
\end{equation}
where $h_n^{\mathrm{CS}}$ denotes the A2G small-scale fading coefficient (Nakagami-$m$), and $n^{\mathrm{CS}}\sim\mathcal{CN}(0,\sigma_{\mathrm{CS}}^{2})$ is the AWGN at the central server. With unit transmit symbol power at the GU, the AF gain $G_{m,n}(k)$ is chosen to satisfy the per-stream transmit power constraint 
\[
    \mathbb{E}\!\left[\,\big|G_{m,n}(k)\,y_n^{\mathrm{UAV}}(k)\big|^{2}\,\right] = \frac{P^{\mathrm{UAV}}}{M_n(k)}.
\]
The resulting gain expression is  
\begin{equation}
    G_{m,n}(k) = 
    \sqrt{
        \frac{\displaystyle \frac{P^{\mathrm{UAV}}}{M_n(k)}}
             {\big|h_{m,n}^{\mathrm{GU}}\big|^{2}\,\big(d^{\mathrm{GU}}_{m,n}(k)\big)^{-2p} + \sigma^{2}_{\mathrm{UAV},n}}
    }.
    \label{eq:af_gain_unit}
\end{equation}

Let $\mathcal{N}_m(k)=\{\,n:\delta_m^n(k)=1\,\}$ be the set of UAVs serving GU $m$ in slot $k$, with cardinality $N_m(k)=|\mathcal{N}_m(k)|$. The data of GU $m$ is split according to $\rho_m^n(k)$ with $\sum_{n\in\mathcal{N}_m(k)}\rho_m^n(k)=1$. The time required for $m^{th}$ GU data transmission over $n^{th}$ UAV is determined as:
\begin{equation}
    T^{m}_{n}(k)=\frac{\rho_m^n(k)\,D_m(k)}{R^{m}_{n}(k)} 
\end{equation}

here the $R^m_n(k)$ is the transmission rate for the whole link given as,
\begin{equation}
    R_{n}^m(k)
= \frac{B_u}{M_n(k)}\;
\log_2\!\left(
1+\frac{ \,|h_{n}^{cu}|^{2}\,|h_{m,n}^{gu}.G_{m,n}(k)|^{2}\,\big(d_{m,n}^{\mathrm{GU}}(k)\big)^{-2p} }
{ \big(G_{m,n}(k)\,|h_{n}^{cu}|\,\sigma_{n}^{\mathrm{uav}}\big)^{2} + \big(\sigma^{cs}\big)^{2} }
\right).
\end{equation}
here $B_u$ denotes the total uplink bandwidth available to each UAV. So, the overall time required for transmission of $m^{th}$ GU data will be, 
\begin{equation}
    T^{m}(k)=\max_{\,n\in\mathcal{N}_m(k)}\,T^{m}_{n}(k)
\end{equation}
The transmission of GU $m$ in slot $k$ is completed at time $T^{m}(k)$, and the 
corresponding completion timestamp is recorded as ${t'}_{k,m}$. This timestamp 
must satisfy 
\[
    {t'}_{k,m} < k\tau,
\]
otherwise, the task will be dropped.

The total energy consumed by the $n^\mathrm{th}$ UAV in receiving data from the 
$m^\mathrm{th}$ GU and subsequently transmitting it to the CS can be expressed as follows:

\begin{align}
    E^{\mathrm{Rx}}_{m,n}(k) &= 
    \frac{
        |h_{m,n}^{\mathrm{gu}}|^{2}
    \big(d_{m,n}^{\mathrm{GU}}(k)\big)^{-2p}
    \cdot \rho_m^n(k) D_m(k)
    }{
        \frac{B_u}{M_n(k)} 
        \log_2 \left( 
            1+ 
            \frac{
                |h_{m,n}^{\mathrm{gu}}|^{2}
                \big(d_{m,n}^{\mathrm{GU}}(k)\big)^{-2p}
            }{
                (\sigma_n^{\mathrm{UAV}})^2
            } 
        \right)
    } \label{eq:Erx}\\[6pt]
    E^{\mathrm{Tx}}_{m,n}(k) &= 
    \frac{P^{\mathrm{UAV}}}{M_n(k)} 
    \cdot 
    \frac{
        \rho_m^n(k) D_m(k)
    }{
        \frac{B_u}{M_n(k)} 
        \log_2 \left( 
            1 + 
            \frac{
                P^{\mathrm{UAV}}
            }{
                (\sigma_n^{\mathrm{UAV}})^2 M_n(k)
            } 
        \right)
    } \label{eq:Etx}\\[6pt]
    E^{\mathrm{Comm}}_n(k) &= 
    \sum_{m=1}^{M} 
    \delta_m^n(k)
    \left[ 
        E^{\mathrm{Rx}}_{m,n}(k) + 
        E^{\mathrm{Tx}}_{m,n}(k) 
    \right] \label{eq:Ecomp}
\end{align}

\section{Problem Formulation}

We jointly optimize the UAV trajectories $s_n^{UAV}(t), \forall n$, the task proportion ratios $\rho$ and the compression factors $d$, which determine the degree of semantic compression applied to the transmitted data. The objective function is composed of two primary metrics. 

\begin{itemize}
    \item \textbf{Average Age of Information (AoI)}: For each GU $m$, the instantaneous AoI in
time slot $k$ is defined as 
$
\Delta_{m}(k) = t'_{m,k} - t_{m,k},$
where $t_{m,k}$ is the generation timestamp and $t'_{m,k}$ is the completion 
timestamp of the data transmission. The average AoI is obtained by averaging first 
over all GUs and then over all time slots: 
\begin{equation}
    \overline{\Delta} = \frac{1}{K} \sum_{k=1}^{K} \frac{1}{M} \sum_{m=1}^{M} \Delta_{m}(k).
\end{equation}

    \item \textbf{Semantic Structural Similarity (SSS)}: which captures both the semantic integrity and perceptual quality of the transmitted data. It is defined as a weighted sum of the cosine semantic similarity between feature vectors $F$ and $\hat{F}$, denoted as $\mathrm{CosSim}(F,\hat{F})$, and the multi-scale structural similarity (MS-SSIM) between the original image $I$ and 
the reconstructed image $\hat{I}$, denoted as $\mathrm{MS\text{-}SSIM}(I,\hat{I})$: 
\begin{equation}
    \mathrm{SSS} = \alpha_s \cdot \mathrm{CosSim}(F,\hat{F}) + (1-\alpha_s) \cdot 
\mathrm{MS\text{-}SSIM}(I,\hat{I}),
\end{equation}

where $0 \leq \alpha_s \leq 1$ is the weighting parameter. We aim to maximize the 
minimum SSS across all GUs and time slots, i.e., $\min_{m,k} \mathrm{SSS}_{m,k}$.
\end{itemize}
Combining these objectives, the overall optimization problem can be formulated as 
\begin{align}
\min_{\mathbf{s}, \boldsymbol{\rho}, \mathbf{d}} \quad 
    & \overline{\Delta} 
    - \beta \cdot 
      \min_{m,k} \mathrm{SSS}_{m,k} \label{eq:objective} \\[4pt]
\text{s.t.} \quad 
    & \textbf{(C1) } 
      \|s_n(t) - s_{n'}(t)\| \geq D_{\min}, 
      \quad \forall n \neq n', t \nonumber \\
    & \textbf{(C2)} 
     \sum_{k=1}^K E^{\mathrm{Comm}}_n(k) + E^{\mathrm{St}}_n(k) \leq E_{\max}, 
      \quad \forall n, k \nonumber \\
    & \textbf{(C3)} 
      \Delta_{m}(k) \leq \frac{1}{\lambda_m}, 
      \quad \forall m, k \nonumber
\end{align}
Here $\beta$ is the weight balancing factor for AoI and SSS objectives, constraint (C1) enforces a minimum separation $D_{\min}$ between any two UAVs to avoid collisions. (C2) limits the total energy used for movement and transmission by each UAV by $E_{\max}$. (C3) ensures there will be no data backlog and processing overhead at the GU. As the formulated problem is non-convex, we employ an RL algorithm to handle the high-dimensional search space and dynamic environments. This system is affected by two types of noise: (i) semantic noise, which arises from the misalignment between the semantic encoder and decoder, and (ii) physical channel noise. To address these, we first develop the semantic encoder and decoder separately, explicitly accounting for semantic loss. The semantic features extracted from this module are then integrated into the overall system, where the RL algorithm is applied for optimal decision-making.

\subsection{ViT-CNN based Prompt aware Encoder-Decoder}

\begin{figure*}[t]
    \centering
    \includegraphics[width=1.0\linewidth]{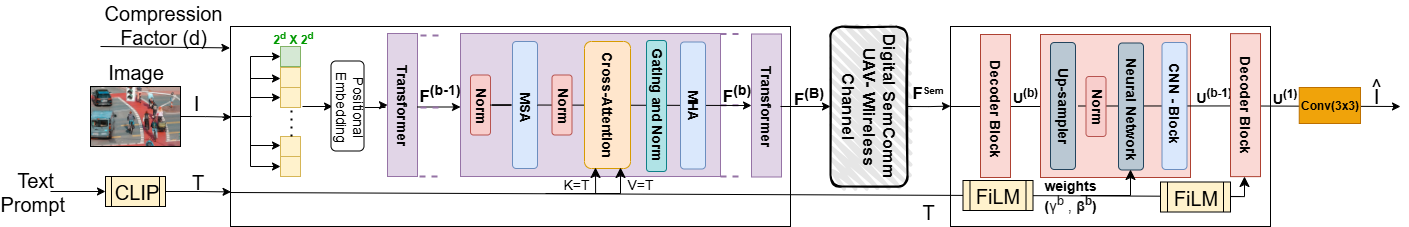} 
    \caption{Encoder - Decoder Architecture}
    \label{fig:ED}
\end{figure*}

We propose a joint semantic communication architecture that integrates a Vision Transformer (ViT)-based encoder, denoted as $V_\theta(\cdot; \theta)$, with a CNN-based decoder, denoted as $C_\phi(\cdot; \phi)$, enhanced with text-prompt guidance for improved context awareness. As shown in Fig.~\ref{fig:ED}, he model takes as input an image
$
I \in \{0,1,\ldots,255\}^{C \times H \times W}
$
and text-prompt tokens
$
T \in \mathbb{R}^{L \times C}
$
obtained from a CLIP\footnote{\href{https://github.com/openai/CLIP}{https://github.com/openai/CLIP}} text encoder, and reconstructs the image at the receiver as
$
\widehat{I} \in \mathbb{R}^{C \times H \times W}.$ The input image is first partitioned into non-overlapping patches of size \(2^{d} \times 2^{d}\), where \(d\) is the compression ratio. Each patch’s \(C\) channels are cascaded into a vector and projected into a token embedding. Formally, for the encoder $V_\theta$, parameterized by $\theta$ (encompassing $\mathbf{W}_e, \mathbf{b}_e$, and all subsequent Transformer block weights), the patch embedding and the Transformer block operations are:
\begin{align}
\mathbf{v}_n &= \mathrm{vec}\big(I[:, r_n{:}r_n + 2^{d} -1, c_n{:}c_n + 2^{d} -1]\big) \in \mathbb{R}^{C \cdot 2^{2d}}, \nonumber \\
\mathbf{x}_n &= \mathbf{W}_e \mathbf{v}_n + \mathbf{b}_e, \quad \mathbf{x}_n \in \mathbb{R}^{C}, \nonumber \\
X^{(0)} &= \{\mathbf{x}_n\}_{n=1}^{N(d)}, \quad N(d) = \frac{HW}{2^{2d}}, \nonumber \\
X^{(b + \frac{1}{3})} &= X^{(b)} + \mathrm{MSA}(\mathrm{LN}(X^{(b)})), \nonumber \\
X^{(b + \frac{2}{3})} &= X^{(b + \frac{1}{3})} + g_b \, \mathrm{CrossAttn}(\mathrm{LN}(X^{(b + \frac{1}{3})}); K=T, V=T), \nonumber \\
X^{(b + 1)} &= X^{(b + \frac{2}{3})} + \mathrm{MLP}(\mathrm{LN}(X^{(b + \frac{2}{3})})),
\end{align}
where \(b = 0, \ldots, B-1\) and \(g_b\) is a learnable gate controlling the strength of text-prompt injection. The output tokens from encoder $V_\theta$ at each block are reshaped into feature maps
\[
F^{(b)} \in \mathbb{C}^{C \times \frac{HW}{2^{2d}}},
\]
which are used for intermediate supervision. We define the final semantic features extracted by the encoder as $F_{sem} = V_\theta(I, T, d; \theta)$, which corresponds to $F^{(B)}$.

The decoder $C_\phi$, parameterized by $\phi$ (encompassing weights for Upsample, Align, CNNBlock, Conv, and the generation of FiLM parameters $\gamma^{(b)}, \beta^{(b)}$), reconstructs \(\widehat{I}\) from the received semantic features, which are based on $\{F^{(b)}\}$ (specifically $F^{(B)}$ at the coarsest level), using a top-down fusion approach. Starting from the coarsest feature \(F^{(B)}\), the decoder $C_\phi$ upsamples and refines the features through convolutional blocks modulated by Feature-wise Linear Modulation (FiLM\footnote{\href{https://github.com/ethanjperez/film}{https://github.com/ethanjperez/film}}) parameters derived from the text prompt. At each decoder stage \(b\), the feature map is updated as:
\begin{align}
U^{(b)} &= \mathrm{Upsample}(U^{(b+1)}) + \mathrm{Align}(F^{(b)}), \nonumber \\
\widetilde{U}^{(b)} &= \gamma^{(b)} \odot \mathrm{LN}(U^{(b)}) + \beta^{(b)},\nonumber \\
U^{(b)} &= \widetilde{U}^{(b)} + \mathrm{CNNBlock}(\widetilde{U}^{(b)}),
\end{align}
where \(\mathrm{Align}(\cdot)\) matches feature scales and channels, and \((\gamma^{(b)}, \beta^{(b)})\) are FiLM parameters computed from the pooled text embedding. The final reconstruction at the finest scale is obtained as
\[
\widehat{I} = \mathrm{Conv}_{3 \times 3}(U^{(1)}).
\]
Formally, we can write $\widehat{I} = C_\phi(F_{sem}, T; \phi)$.

To encourage semantic fidelity, the decoder $C_\phi$ produces intermediate reconstructions at each stage:
\[
\widehat{I}^{(b)} = \mathrm{Conv}_{3 \times 3}(U^{(b)}), \quad b = 1, \ldots, B,
\]
enabling progressive supervision across multiple scales. The model is trained end-to-end by minimizing the multi-scale MS-SSIM loss:
\begin{align}
\mathcal{L} = \sum_{b=1}^B \lambda_b \left(1 - \mathrm{MS\text{-}SSIM}(I, \widehat{I}^{(b)})\right),
\end{align}
where \(\lambda_b\) weights the contribution of each scale. This loss promotes perceptual quality and robustness against channel distortions by enforcing reconstructability of semantic features produced by $V_\theta$ at each encoder depth.

In this design, the ViT-CNN prompt-aware encoder-decoder jointly leverages a cascaded patch embedding Vision Transformer enhanced with text-prompt cross-attention, denoted as $V_\theta(\cdot; \theta)$, and a CNN decoder conditioned on the same prompt via FiLM modulation, denoted as $C_\phi(\cdot; \phi)$. Intermediate reconstructions after each encoder block provide strong multi-scale supervision, resulting in context-aware semantic features \(F \in \mathbb{C}^{C \times \frac{HW}{2^{2d}}}\) that are compact and robust for wireless semantic transmission.

\subsection{MDP Formulation and Algorithm}

In the proposed system, UAVs collaboratively optimize their movement direction, travel distance, task proportion ratio, and compression factor. These actions directly impact the environment by altering interference levels and resource allocation, thereby influencing system performance. The environment evolves stochastically, with state transitions determined by current conditions and joint UAV actions. This dynamic interaction is modeled as a multi-agent Markov Decision Process (MDP):
\[
\langle \mathcal{N}, \mathcal{S}, \{\mathcal{A}_n\}_{n \in \mathcal{N}}, \{\mathcal{R}_n\}_{n \in \mathcal{N}}, \gamma \rangle,
\]
where $\mathcal{N}$ is the set of UAVs, $\mathcal{S}$ the global state space, $\mathcal{A}_n$ the action space, $\mathcal{R}_n$ the reward function of UAV $n$, and $\gamma \in [0,1]$ the discount factor.

\begin{itemize}
    \item \textbf{Action Space ($A_n$):} Action governs the UAVs’ trajectory, task proportion ratio, and semantic data compression level.
Each UAV $n \in \mathcal{N}$ selects an action at time $t$ defined as  
\[
a_n(t) = \left\{ \hat{\omega}_n(t), l_n^{\mathrm{UAV}}(t), \rho_m^n(k), d(k) \right\},
\]  
where $\hat{\omega}_n(t)$ denotes the movement direction, $l_n^{\mathrm{UAV}}(t)$ the travel distance, $\rho_m^n(k)$ the task proportion allocated to GU $m$, and $d(k)$ the data compression ratio at decision frame $k = \left\lfloor \frac{t}{\tau} \right\rfloor$.

\item \textbf{State Space (S):}  State captures the network-wide mobility, data request, energy, and wireless channel dynamics at time $t$.
The system state at time $t$ is given by  
\[
s(t) = \Big\{
\underbrace{ \left\{ \mathbf{s}_n^{\mathrm{UAV}}(t) \right\}_{n \in \mathcal{N}} }_{\text{UAV positions}}, \quad
\underbrace{ \left\{ \mathbf{s}_m^{\mathrm{GU}}(t), v_{m,k}^{\mathrm{GU}}, D_m(k), \lambda_m \right\}_{m \in \mathcal{M}} }_{\text{GU position, velocity, data, arrival rate}},
\]
\[
\underbrace{ \left\{ E_n^{\mathrm{St}}(k), E_n^{\mathrm{Comm}}(k) \right\}_{n \in \mathcal{N}} }_{\text{UAV energy status}}, \quad
\underbrace{ \left\{ h_{m,n}^{\mathrm{GU}}, h_n^{\mathrm{UC}} \right\}_{m \in \mathcal{M}, n \in \mathcal{N}} }_{\text{GU–UAV and UAV–CS channel states}} 
\Big\}.
\]

\item Reward $R_n(t)$: The reward for UAV $n$ at time $t$, denoted by $R_n(t)$, comprises two parts: a system-wide reward shared by all UAVs at mission completion, and penalties for constraint violations. The system reward will be the overall system cost Eq.~\ref{eq:objective}, defined as  
\[
R_{\text{sys}} =  \beta \cdot \min_{m,k} \mathrm{SSS}_{m,k} - \overline{\Delta} ,
\]

Penalties are applied if constraints are violated, $\eta_1$ will be applied if collision avoidance constraints are not satisfied, $\eta_2$ if UAV energy constraints are violated, $\eta_3 \sum_{m} \delta_m^{(n)}(k) \cdot \frac{1}{\lambda_m}$ penalize for AoI voilation for $m^{th}$ GU in time slot $k$.

Hence, the total reward for UAV $n$ at time $t$ is given by  
\begin{align}
R_n(t) =\; & R_{\text{sys}} 
- \eta_1 \cdot \mathbf{1}_{\text{collision}} - \eta_2 \cdot \mathbf{1}_{\text{energy}} \nonumber \\[-6pt]
& - \eta_3 \sum_{m} \delta_m^{(n)}(k) \cdot \frac{1}{\lambda_m},
\end{align}

where $\mathbf{1}_{\text{collision}}$ and $\mathbf{1}_{\text{energy}}$ are indicator functions equal to 1 when collision or energy constraints are violated, respectively.

\end{itemize}

\textbf{Truncated Quantile Critic (TQC) Algorithm:} For learning, we employ the TQC algorithm for its robust ability to handle continuous action spaces, mitigate overestimation bias via distributional learning and quantile truncation, and stabilize training through critic ensembling and entropy regularization \cite{kuznetsov2023adapting}. The comprehensive training process for our multi-UAV system using TQC is detailed in Algorithm ~\ref{alg:tqc_semantic_uav}. The core of this learning process involves iteratively updating the actor and critic networks based on experiences sampled from a replay buffer $\mathcal{D}$.

At each time step, the agent interacts with the environment by executing actions, which include UAV mobility, resource allocation, and the critical semantic data compression ratio ($d(k)$). The resulting image reconstruction quality, determined by our semantic encoder-decoder, contributes to the immediate reward $R_t$. The reward formulation incorporates the multi-scale MS-SSIM loss for semantic fidelity, defined as:
\begin{align}
\mathcal{L}_{\text{MS-SSIM}}(I_t, \widehat{I}_t) = \sum_{b=1}^B \lambda_b \left(1 - \mathrm{MS\text{-}SSIM}(I_t, \widehat{I}_t^{(b)})\right)
\end{align}
alongside considerations for communication efficiency, energy consumption, and AoI. Each observed transition $(s_t, a_t, R_t, s_{t+1})$ is stored in the replay buffer $\mathcal{D}$ to facilitate off-policy learning.

During the learning phase, mini-batches of transitions are sampled from $\mathcal{D}$ to update the network parameters. The ensemble of $K$ critic networks, parameterized by $\psi_j$, are updated by minimizing the quantile Huber loss $\mathcal{L}_{\kappa}$ against a common target value $y_t$. This target is derived from the Bellman equation, incorporating a bias-reduced estimate of the next state-action value obtained from the target networks:
\begin{align}
y_t = R_t + \gamma \left( \bar{Q}_{\text{trunc}}(s_{t+1}, a'_{t+1}) - \alpha \log \pi(a'_{t+1}|s_{t+1}) \right)
\end{align}
where $a'_{t+1}$ is sampled from the target policy $\bar{\pi}(\cdot|s_{t+1})$. The term $\bar{Q}_{\text{trunc}}(s_{t+1}, a'_{t+1})$ represents the truncated mean of the quantiles predicted by the ensemble of target critic networks. Specifically, the $KN$ quantiles $\{\bar{Z}_j(s_{t+1}, a'_{t+1})\}_{j=1 \dots K, i=1 \dots N}$ are combined and sorted as $\tilde{z}_1 \le \dots \le \tilde{z}_{KN}$. The truncated mean is then computed by discarding the top $d_{\text{trunc}}$ quantiles:
\begin{align}
\bar{Q}_{\text{trunc}}(s_{t+1}, a'_{t+1}) = \frac{1}{KN-d_{\text{trunc}}} \sum_{k=1}^{KN-d_{\text{trunc}}} \tilde{z}_k
\end{align}
The loss for each critic $Q_j$ is then defined as:
\begin{align}
\mathcal{L}_{Q_j}(\psi_j) = \mathbb{E} \left[ \sum_{i=1}^N \mathcal{L}_{\kappa}(Z_i(s_t, a_t; \psi_j) - y_t) \right]
\end{align}

The actor network, parameterized by $\phi$, is updated via policy gradients to maximize the expected value of actions. The objective function for the actor leverages the robust Q-value estimate derived from the online critics:
\begin{align}
\mathcal{L}_{\pi}(\phi) = \mathbb{E} \left[ \alpha \log \pi(a_t|s_t) - \text{TruncatedQValue}(s_t, a_t) \right]
\end{align}
Here, $\text{TruncatedQValue}(s_t, a_t)$ is computed similarly to $\bar{Q}_{\text{trunc}}$, but using the quantiles from the ensemble of online critic networks for the current state-action pair $(s_t, a_t)$. To stabilize training and prevent oscillations, target networks for both actor and critics are updated periodically using soft averaging:
\begin{align}
\bar{\theta} \leftarrow \tau \theta + (1 - \tau) \bar{\theta}
\end{align}
where $\theta$ represents the parameters of the online networks (actor $\phi$ or critics $\psi_j$) and $\bar{\theta}$ represents their target counterparts.

This structured approach ensures that the learning process for UAV policies is stable, efficient, and directly integrates the quality of semantic information reconstruction into the overall optimization objective.

\begin{algorithm}[t]
\small
\caption{ Semantic-Aware TQC for Multi-UAV Systems}
\label{alg:tqc_semantic_uav}
\begin{algorithmic}[1]
\REQUIRE Semantic Encoder $V_\theta$, Decoder $C_\phi$, CLIP; TQC hyperparameters; Semantic hyperparameters.

\STATE \textbf{Initialize:} Actor $\pi(\phi)$, Critic $Q_j(\psi_j)$ and their target networks; Replay buffer $\mathcal{D}$.

\FOR{each training episode}
    \STATE Observe initial global state $s_t$.
    \FOR{each time step $t$}
        \STATE Obtain image $I_t$ and prompt $T_t$.
        \STATE UAV selects action $a_n(t)$ (including compression ratio $d(k)$) based on $\pi_n(\cdot|s(t))$.
        \STATE Perform SemComm: $F_t = V_\theta(I_t, T_t, d(k))$; Transfer $F_t$ to channel; $\widehat{I}_t = C_\phi(F_t^{\text{rec}}, T_t)$.
        \STATE Compute reward $R_t$ (incorporating semantic quality $\mathcal{L}_{\text{MS-SSIM}}$, communication, energy, AoI).
        \STATE Observe $s_{t+1}$. Store $(s_t, a_t, R_t, s_{t+1})$ in $\mathcal{D}$.
    \ENDFOR

    \FOR{each training iteration}
        \STATE Sample mini-batch from $\mathcal{D}$.
        \STATE Compute target $y_t$ for critic update Eq (25).
        \STATE Update critic parameters $\psi_j$ by minimizing Eq. (27).
        \STATE Update actor parameters $\phi$ by minimizing Eq. (28).
        \STATE Soft update target networks ($\bar{\phi}, \bar{\psi}_j$).
    \ENDFOR
\ENDFOR

\RETURN Optimized Actor $\pi(\phi)$ and Critic $Q_j(\psi_j)$ networks.
\end{algorithmic}
\end{algorithm}

\section{Simulation Results and Discussion}
In this section, we rigorously evaluate the performance of our proposed Deep Semantic Communication (DSC)-UAV model, trained with a Task Quality Criterion (TQC) algorithm, against several established and custom baseline approaches. Our objective is to demonstrate the superior efficiency and robustness of semantic-aware communication and UAV-aided data relaying in dynamic environments.

\subsection{Simulation Setup}
We consider a two-dimensional operational region spanning $[-1000, 1000] \times [-1000, 1000]$ m. The central server is statically positioned at the origin $(0,0,0)$. Our simulations involve $M=20$ GUs and $N=2$ Unmanned Aerial Vehicles (UAVs) acting as mobile relay nodes. Each UAV flies at an altitude dynamically varying within the range of $100$ m to $150$ m, balancing regulatory constraints and optimal coverage in urban environments. Initial GU positions are drawn uniformly at random over the area and follow a random waypoint mobility model throughout the mission.

The simulation environment was developed using MATLAB R2023b, leveraging the Mapping Toolbox \cite{MATLAB_Mapping_Toolbox_Ref}, Deep Learning Toolbox \cite{MATLAB_DL_Toolbox_Ref}, and Wireless Communication Toolbox \cite{MATLAB_WC_Toolbox_Ref}. The semantic ViT-CNN encoder-decoder model was implemented in TensorFlow/Keras \cite{TensorFlow_Keras_Ref} and integrated into MATLAB through the `pyrun` interface \cite{MATLAB_pyrun_Ref}, enabling dynamic semantic processing during simulation. For training the semantic encoder-decoder, we used a curated subset of the KITTI raw dataset \cite{Geiger2013IJRR}, which consists of $1242 \times 375$ RGB images depicting real-world road scenes under diverse lighting and environmental conditions. For testing, a disjoint set of traffic images from an online source \cite{Shutterstock_Germany_Road_Traffic} was used. These were resized and normalized to ensure compatibility with the training data distribution. The mission duration is set to $1000$ seconds to capture long-term system behavior, including UAV scheduling convergence, semantic performance degradation, and GU coverage fairness. Key simulation parameters are summarized in Table~\ref{tab:simulation_parameters}. 


\begin{table}[!t]
\centering
\caption{Key Simulation Parameters}
\label{tab:simulation_parameters}
\begin{tabular}{|l|l|l|}
\toprule
\textbf{Symbol} & \textbf{Description} & \textbf{Value} \\
\midrule
$A$                        & Region of operation                  & $2 \times 2$ $km^2$ \\
$M$                        & Number of GUs         & 20 \\
$N$                        & Number of UAVs                       & 2 \\
$Z^{\mathrm{UAV}}$         & UAV altitude                         & [100,150] m \\
$v^{\mathrm{GU}}$          & GU speed range                       & [0.3, 1.5] m/s \\
$v_{\max}^{\mathrm{UAV}}$  & Max UAV speed                        & 15 m/s (54 km/h) \\
$\lambda_m$               & GU image arrival rate                & [0.05, 0.2] images/s \\
$d$                        & Compression factor         &  [1,4] \\
$T$                        & Mission duration                     & 1000 s \\
$\tau$                     & Time slot duration                   & 5 s \\
$\alpha_{\mathrm{hover}}$  & UAV hovering power coefficient                   & 120 W \\
$h$                        & Channel model                        & Nakagami-$m$ ($m = 2$) \\
$f_c$                      & Carrier frequency                    & 2.4 GHz \\
$p$                        & Path loss exponent                   & 2.7 \\
$P^{\mathrm{UAV}}$         & UAV transmit power                   & 200 mW (23 dBm) \\
$(\sigma_n^{\mathrm{UAV}})^2$ & UAV receiver noise power           & -105 dBm \\
$(\sigma^{\mathrm{CS}})^2$ & Central server noise power           & -105 dBm \\
$B_u$                      &  Uplink bandwidth               & 10 MHz \\
$C \times H \times W$      & Image resolution (RGB)               & $3 \times 375 \times 1242$ \\
$\alpha_r$                & UAV coverage angle             & $60^\circ$ \\
\bottomrule
\end{tabular}
\end{table}

\begin{figure}[!ht]
    \centering
    \includegraphics[width=1\linewidth]{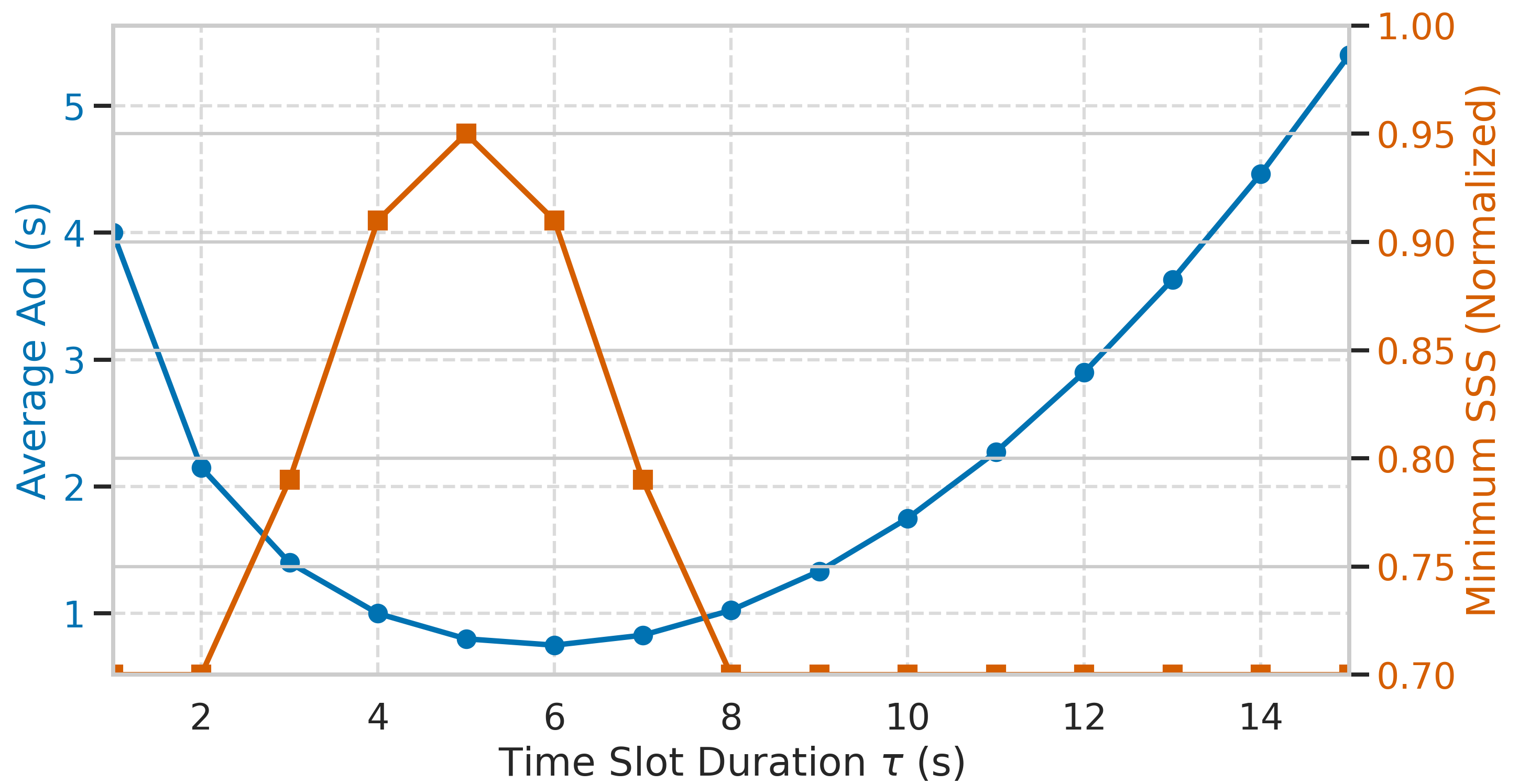} 
    \caption{Effect of time slot duration \( \tau \) on system performance}
    \label{fig:tau_effect_on_system_performance}
\end{figure}

We analyze the role of time slot duration \( \tau \) in balancing timeliness and semantic accuracy. Small \( \tau \) values allow more frequent updates, enhancing system responsiveness. However, the limited time per slot leads to incomplete transmissions, causing task drops and increased AoI. Moreover, to fit data into shorter slots, higher compression is required, which degrades semantic quality (SSS). On the other hand, large \( \tau \) values provide ample time for transmission, reducing the need for compression and improving semantic fidelity. Yet, longer slots result in idle periods after early task completion, leading to underutilized resources and fewer updates. Additionally, longer intervals allow user mobility cause channel phase variations as frequency drift component becomes dominant, introducing semantic mismatches. Considering ground user mobility and arrival patterns, we find that \( \tau = 5 \)~s offers a suitable balance, as supported by the trends observed in Fig.~\ref{fig:tau_effect_on_system_performance}.

\subsection{Performance Comparison with Baselines}

Table~\ref{tab:comparison_snr} summarizes the performance of our proposed DSC+TQC model compared to four baselines across various SNR values. Our approach consistently outperforms others in both Average AoI and Minimum SSS. In the practically relevant SNR range of 5-10~dB, DSC-UAV+TQC achieves approximately 14\% lower AoI and 22\% higher SSS relative to the purely digital communication baseline (D+TQC). This improvement arises from the effective integration of semantic and digital communication, which enables efficient compression and prioritization of task-relevant data while maintaining physical layer robustness. Compared to the purely semantic approach (SC+TQC), our model benefits from a strong digital transmission backbone, ensuring reliable communication even under challenging channel conditions. Moreover, against advanced RL-based baselines such as DSC+SAC and DSC+TD3, the use of the Truncated Quantile Critic (TQC) algorithm yields superior UAV control and resource allocation, further enhancing both AoI and semantic fidelity.

\begin{table}[t]

\centering
\caption{Comparison of Average AoI and Minimum SSS over SNR for Different Algorithms}
\label{tab:comparison_snr}
\setlength{\tabcolsep}{4pt} 
\renewcommand{\arraystretch}{1.2} 
\rowcolors{2}{gray!10}{white} 
\begin{tabular}{|c|cc|cc|cc|cc|cc|}
\hline
\rowcolor{gray!25}
\textbf{SNR} 
& \multicolumn{2}{c|}{\textbf{DSC+TQC}} 
& \multicolumn{2}{c|}{\textbf{D+TQC}} 
& \multicolumn{2}{c|}{\textbf{SC+TQC}} 
& \multicolumn{2}{c|}{\textbf{DSC+SAC}} 
& \multicolumn{2}{c|}{\textbf{DSC+TD3}} \\ 
\rowcolor{gray!25}
\textbf{dB} & \textbf{AoI} & \textbf{SSS} & \textbf{AoI} & \textbf{SSS} & \textbf{AoI} & \textbf{SSS} & \textbf{AoI} & \textbf{SSS} & \textbf{AoI} & \textbf{SSS} \\ \hline

0  & 4.7  & \textbf{0.76} & 5.3  & 0.64 & 5.6  & 0.72 & 5.0  & 0.70 & 5.1  & 0.71 \\ 
5  & 4.0  & \textbf{0.83} & 4.8  & 0.69 & 5.1  & 0.78 & 4.4  & 0.76 & 4.5  & 0.77 \\ 
10 & \textbf{3.4}  & \textbf{0.91} & 3.9  & 0.75 & 4.1  & 0.88 & 3.7  & 0.85 & 3.6  & 0.87 \\ 
15 & \textbf{3.1}  & \textbf{0.93} & 3.6  & 0.78 & 3.8  & 0.89 & 3.5  & 0.87 & 3.4  & 0.89 \\ 
20 & \textbf{2.9}  & \textbf{0.94} & 3.4  & 0.80 & 3.6  & 0.90 & 3.3  & 0.89 & 3.2  & 0.90 \\ \hline

\end{tabular}
\end{table}

\subsection{Impact of Semantic Feature Length and Modulation Order}


\begin{figure}[!t]
    \centering 
    \begin{subfigure}{0.8\linewidth}
        \centering
        \includegraphics[width=\linewidth]{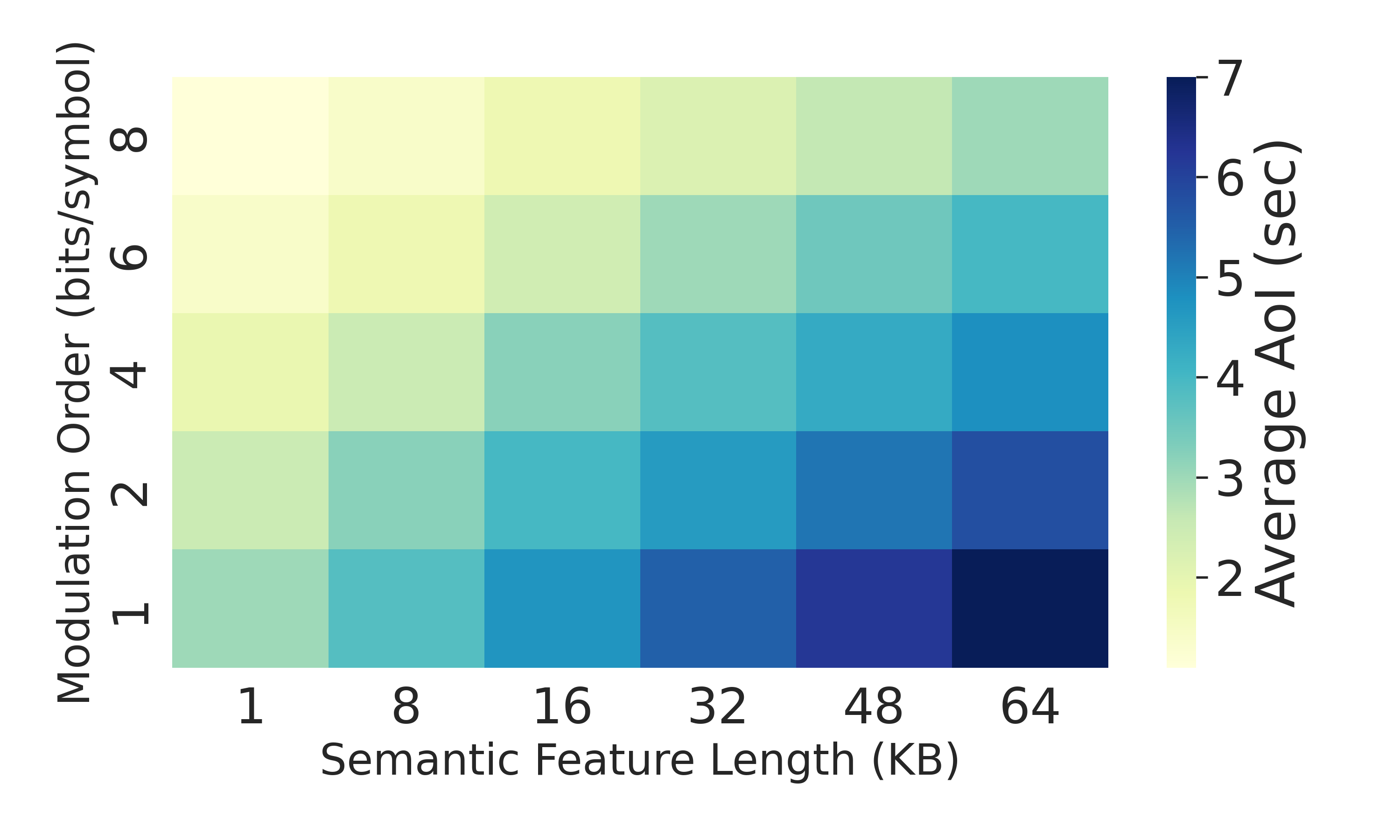}
        \caption{Avg AoI}
        \label{fig:aoi_heatmap}
    \end{subfigure}
    \begin{subfigure}{0.8\linewidth}
        \centering
        \includegraphics[width=\linewidth]{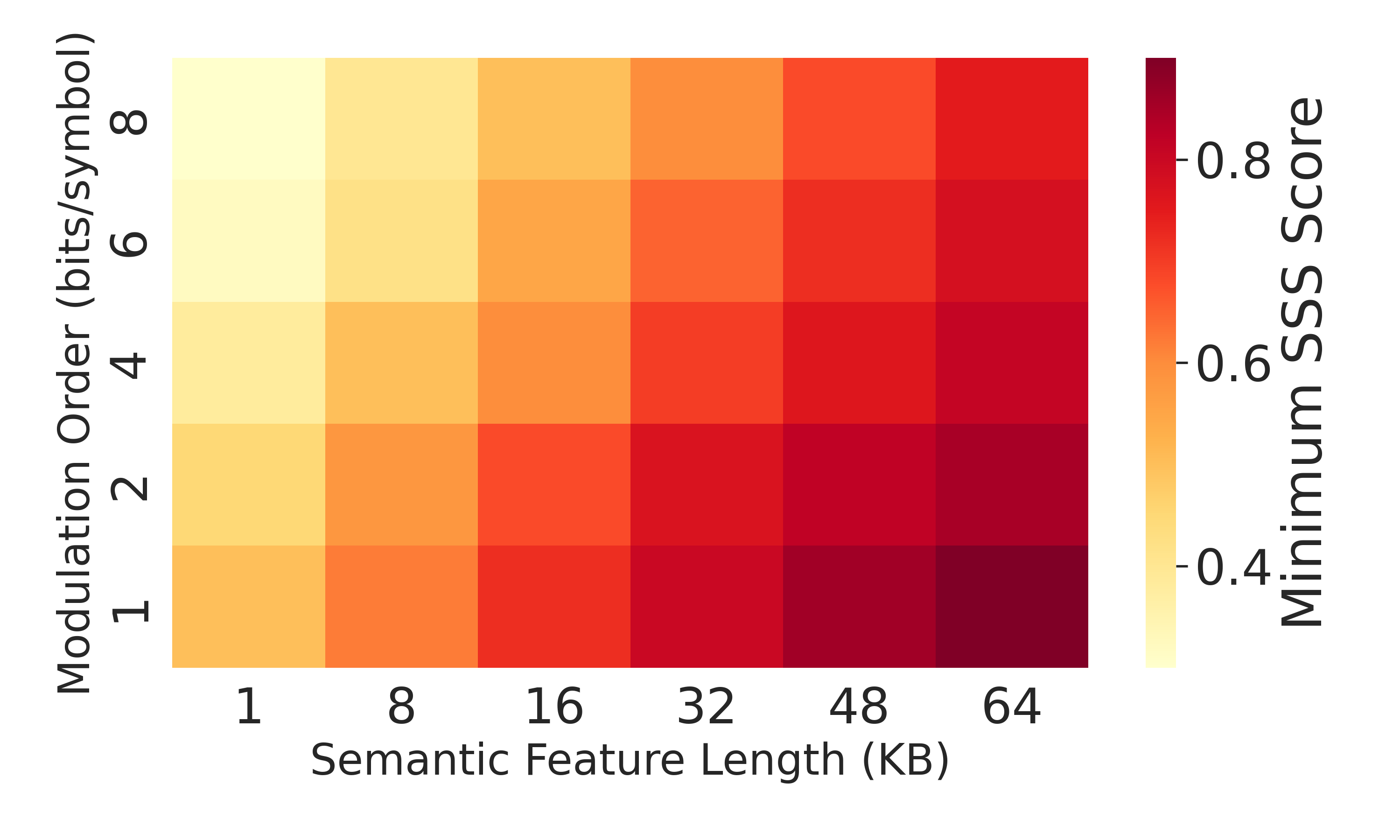}
        \caption{min SSS}
        \label{fig:sss_heatmap}
    \end{subfigure}
    \caption{Effect of semantic feature length and modulation order}
    \label{fig:heatmaps}
\end{figure}

The heatmaps in Fig.~\ref{fig:aoi_heatmap} and Fig.~\ref{fig:sss_heatmap} illustrate the interplay between semantic feature length, modulation order, and system performance metrics— Average AoI and minimum SSS in the whole mission over all GUs. As seen in Fig.~\ref{fig:aoi_heatmap}, AoI generally increases with semantic feature length due to the longer transmission times required for larger encoded data. Conversely, higher modulation orders reduce AoI by enabling faster data transmission through increased bits per symbol. At very low semantic feature lengths, the AoI also decreases with higher modulation since the data packets are smaller, further reducing delay.

Fig.~\ref{fig:sss_heatmap} demonstrates that SSS improves with increased semantic feature length at low modulation orders, reflecting richer semantic information and better reconstruction quality. However, increasing modulation order at high semantic lengths causes a decline in SSS, attributed to higher symbol detection errors in higher-order QAM. Additionally, low semantic feature lengths combined with high modulation orders lead to further degradation in SSS due to the amplified impact of channel noise on already compressed representations. These results highlight a critical tradeoff between transmission delay and semantic fidelity, emphasizing the need to balance semantic compression and modulation level for optimal performance.

\subsection{Prompt-Aware Semantic Transmission under Noise}

\begin{figure}[!ht]
    \centering
    \includegraphics[width=0.98\linewidth]{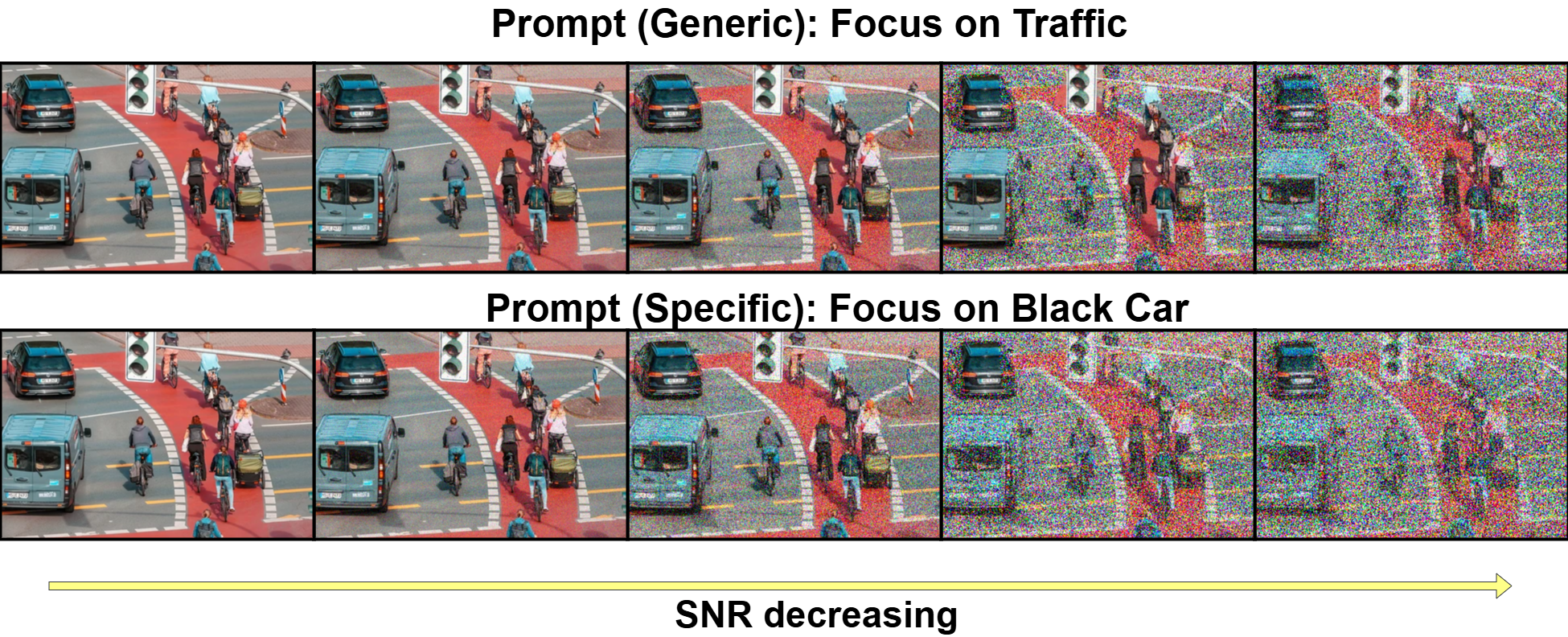} 
    \caption{Prompt-specific vs. generic decoding under low SNR}
    \label{fig:prompt_case}
\end{figure}

To assess the behavior of the proposed prompt-aware encoder-decoder, we conducted two case studies under gradually decreasing SNR:

\textbf{Case 1: Generic Prompt} – "Analyze the traffic scene."  
As shown in Figure~\ref{fig:prompt_case}, the model initially captures various elements such as traffic lights, cars, and cyclists. However, with decreasing SNR, the reconstructed images gradually lose detail, and semantic understanding degrades significantly. At extremely low SNR, no meaningful object can be distinguished, highlighting the sensitivity of generic prompts under harsh conditions.

\textbf{Case 2: Specific Prompt} – "Focus on the black car."  
In this case, even as the SNR drops, the model successfully retains the representation of the black car, while ignoring irrelevant background features. This confirms the effectiveness of task-oriented prompts in guiding semantic compression and reconstruction. The attention focus remains sharper, and the system better resists noise perturbations. More specific prompts result in better alignment with the communication objective, even in noisy environments.

\section{Conclusion}

We presented DSC-UAV, a prompt-aware semantic communication framework for UAV networks operating under bandwidth constraints. By integrating context-driven encoding, adaptive UAV relaying, and TQC-aided mobility optimization, our model ensures efficient and relevant data transmission. Simulation results show a 14\% reduction in Age of Information (AoI) and a 22\% improvement in Semantic Structural Similarity (SSS), demonstrating both timely delivery and efficient semantic compression. These gains confirm that DSC-UAV performs effectively under bandwidth constraints by focusing on context-relevant content and minimizing transmission delays. The proposed framework thus offers a resilient, information-centric solution for UAV communication in dynamic and resource-limited environments.

\bibliographystyle{IEEEtran}
\bibliography{ref}

\end{document}